\documentclass[%
 reprint,
 amsmath,amssymb,prd,twocolumn,superscriptaddress,nofootinbib,
 nolongbibliography
]{revtex4-2}

\usepackage{graphicx}
\usepackage{dcolumn}
\usepackage{bm}
\usepackage{dcolumn}
\usepackage{xcolor}

\usepackage{eso-pic}
\newcommand{\reportnum}[2]{
  \AddToShipoutPictureBG*{%
    \AtPageUpperLeft{%
      \hspace{0.75\paperwidth}%
      \raisebox{#1\baselineskip}{%
        \makebox[0pt][l]{\textnormal{#2}}
  }}}%
}

\begin{document}

\reportnum{-3}{CERN-TH-2023-181}

\title{\boldmath Third order correction to semileptonic $b\to u$ decay: fermionic contributions}

\author{Matteo Fael}
\email{matteo.fael@cern.ch}
\affiliation{Theoretical Physics Department, CERN, 1211 Geneva, Switzerland}
\author{Johann Usovitsch}%
\email{johann.usovitsch@cern.ch}
\affiliation{Theoretical Physics Department, CERN, 1211 Geneva, Switzerland}

\date{\today}

\begin{abstract}
We present the QCD corrections of order $\alpha_s^3$ to the decay rate of $b \to u \ell \bar \nu_\ell$, with $\ell = e,\mu$, originating from 
diagrams with closed fermion loops and neglecting the mass of the up quark. 
Our calculation relies on integration-by-parts reduction of Feynman integrals with one propagator raised to a symbolic power in \texttt{Kira}
and the numerical evaluation of master integrals with \texttt{AMFlow}.
This allows us to obtain results for the fermionic contributions to the total semileptonic 
rate with an accuracy of more than thirty digits. 
\end{abstract}

\preprint{CERN-TH-2023-181}

\maketitle

\section{Introduction}
The inclusive $B$-meson decay $B\to X_u \ell \bar \nu_\ell$, with $\ell = e,\mu$,
has a pivotal role in the extraction of the Cabbibo–Kobayashi–Maskawa matrix element $|V_{ub}|$
and in global fits of the unitarity triangle within the Standard Model~\cite{UTfit:2006vpt,UTfit:2022hsi}.

Because of experimental cuts applied to semileptonic $b \to u$ decay to
suppress the $b\to c$ contamination, the theoretical description of the
differential rate for inclusive $B\to X_u \ell \bar \nu_\ell$ 
is based on a non-local Operator Product Expansion (OPE)~\cite{Neubert:1993ch,Neubert:1993um,Bigi:1993ex}.
Perturbative coefficients in the OPE are convoluted with non-perturbative shape functions. 
Their exact form cannot be calculated from first principles
so different parametrizations exist~\cite{Lange:2005yw,Andersen:2005mj,Aglietti:2007ik,Gambino:2007rp,Bernlochner:2020jlt} 
which lead to $|V_{ub}|$ determinations
with uncertainties of about $4\%$~\cite{HeavyFlavorAveragingGroup:2022wzx}.

In this paper we consider the total $B\to X_u \ell \bar \nu_\ell$ decay rate 
which is cleaner from the theoretical point of view because it does not involve
shape functions. 
The total rate is described by the Heavy Quark Expansion (HQE)~\cite{Manohar:1993qn,Blok:1993va,Bigi:1993fe}, a local OPE in inverse powers of the bottom quark mass, which has been successfully
applied to semileptonic $b \to c$ decays and the extraction of $|V_{cb}|$~\cite{Bordone:2021oof,Bernlochner:2022ucr}.

The Belle II Collaboration~\cite{lucao} has recently presented 
a preliminary measurement of the
ratio $\Gamma(B \to X_u \ell \bar \nu_\ell)/\Gamma(B \to X_c \ell \bar \nu_\ell)$
from which it is possible to extract $|V_{ub}/V_{cb}|$ given a prediction for the phase-space ratio~\cite{Gambino:2001ew,Bobeth:2003at,Gambino:2008fj}
\begin{equation}
    C = \left\vert \frac{V_{ub}}{V_{cb}} \right\vert^2
    \frac{\Gamma(B \to X_c \ell \bar \nu_\ell)}{\Gamma(B \to X_u \ell \bar \nu_\ell)},
\end{equation}
which is also employed to normalize the branching ratio of radiative decays ($B \to X_{s} \gamma$) 
and rare semileptonic decays ($B \to X_{s} \ell^+ \ell^-$).
The ratio $C$ is determined using the HQE together with measurements 
of the $B \to X_c \ell \bar \nu_\ell$ decay spectra.
The current estimate $C= 0.568 \pm 0.007 \pm 0.010$~\cite{Alberti:2014yda} has a 2.1\% 
uncertainty that will be comparable with the future experimental error 
on $B \to X_s \gamma$, of about 2.6\%, achievable with the full Belle II dataset~\cite{Belle-II:2018jsg}. 
Also for the theoretical prediction of the  
$B \to X_{s} \ell^+ \ell^-$ branching fraction, the ratio $C$ is a significant source of uncertainty~\cite{Huber:2005ig,Huber:2019iqf,Huber:2020vup}.

In the free quark approximation the total rate of $ b\to u \ell \bar \nu_\ell$
has been calculated up to $O(\alpha_s^2)$ in Refs.~\cite{vanRitbergen:1999gs,Pak:2008qt,Pak:2008cp}.
The third order correction has been estimated in Refs.~\cite{Fael:2020tow,Fael:2022frj} 
(see also Ref.~\cite{Czakon:2021ybq})
by performing an asymptotic expansion for $\delta = 1-m_c/m_b \to 0$, i.e.\
in the limit $m_b \simeq m_c$.
The series in $\delta$ exhibits a fast convergence and allows to obtain
an accurate result for $b \to c$ decay at the physical value of the charm mass.
Moreover the expansion shows a good convergence even for $\delta \to 1$, 
corresponding to a massless final-state quark. 
This allows to make an estimate of the charmless decay $b\to u$, however with a 10\% uncertainty on the 
$\alpha_s^3$ correction due to the extrapolation to massless quark.
This prediction has been recently confirmed by an independent calculation 
performed in the leading-color approximation~\cite{Chen:2023dsi}
with an uncertainty of about 5\% in the on-shell scheme,
a factor of two improvement compared to Ref.~\cite{Fael:2020tow}.
This translates into a systematic $0.5\%$ uncertainty on the total semileptonic rate.

In the present paper, we take a first step towards the improvement of the 
prediction for the total rate of $b \to u \ell \bar \nu_\ell$.
We present the calculation of the fermionic contributions at order $\alpha_s^3$, i.e.\ 
the subset of five-loop diagrams containing closed fermion loops. 
The evaluation of the remaining part, the bosonic corrections, 
is underway and it will be presented in a future publication.
In Sec.~\ref{sec:methods} we describe the methods used for the calculation 
of five-loop diagrams, in particular how we perform the integration-by-parts (IBP)
reduction and the numerical evaluation of the master integrals.
We present our results in Sec.~\ref{sec:results} together with a comparison 
with previous calculations. We conclude in Sec.~\ref{sec:conc}.

\section{Methods}
\label{sec:methods}

\begin{figure}
    \centering
    \begin{tabular}{cc}
    \includegraphics[width=0.23\textwidth]{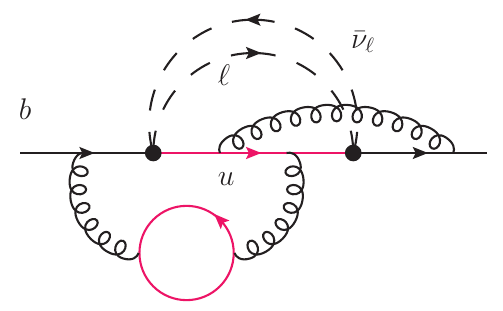} &
    \includegraphics[width=0.23\textwidth]{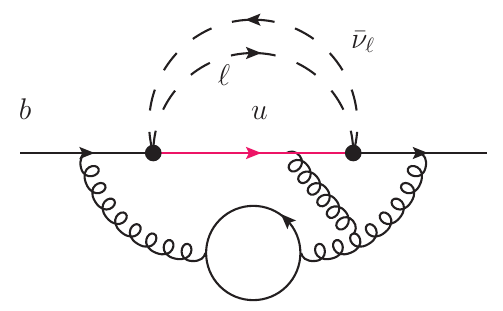}\\
      (a)   &  (b) \\[5pt]
    \includegraphics[width=0.23\textwidth]{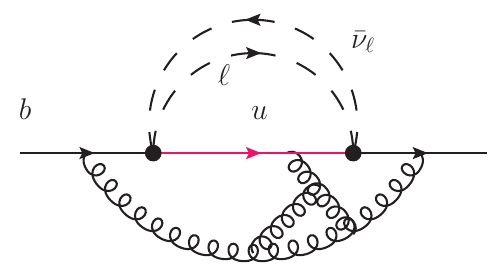} &
    \includegraphics[width=0.23\textwidth]{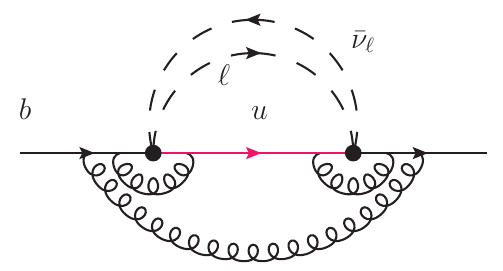}\\
      (c)   &  (d) \\
    \end{tabular}
    \caption{Five-loop diagrams contributing to the $\alpha_s^3$ correction
    to $b \to u \ell \bar \nu_\ell$.
    Sample of fermionic ({a,b}) and bosonic ({c,d}) contributions.
    Lepton and neutrino are shown with dashed lines, black and red solid lines represent 
    the bottom and up quark.
    The effective vertex is shown by a dot.
    }
    \label{fig:dias}
\end{figure}

We calculate higher order QCD corrections to the decay rate by employing the 
optical theorem and considering the imaginary part of self-energy diagrams 
like those shown in Fig.~\ref{fig:dias}. 
At order $\alpha_s^3$ we compute diagrams up to five loops.
The Feynman diagrams contain a neutrino, a charged lepton 
and an up quark as internal particles, which are all considered massless.
Only the bottom quark is taken as massive and we normalize its mass to unity for simplicity.
The Feynman integrals depend only on the dimensional regularization parameter
$d = 4-2\epsilon$.
The weak decay mediated by the $W$ boson is treated with an effective four-fermion interaction,
shown with black dots in Fig.~\ref{fig:dias}. 

In this paper we concentrate on the subset of
gauge-invariant diagrams that contain at least one closed fermion loop,
where the internal quark can be massless, $u,d,s,c$, or massive, the bottom quark [see for instance Figs.~\ref{fig:dias}(a) and~\ref{fig:dias}(b)].
We ignore the finite charm-mass effects.

For the computation of the bottom self-energy diagrams we use \texttt{qgraf}~\cite{Nogueira:1991ex} to generate the amplitude. 
We use the program \texttt{tapir}~\cite{Gerlach:2022qnc} 
for identification, manipulation and minimization of Feynman integral families.
With \texttt{exp}~\cite{Harlander:1998cmq,Seidensticker:1999bb} we generate a 
\texttt{FORM}~\cite{Vermaseren:2000nd,Kuipers:2012rf,Ruijl:2017dtg} code to 
perform the Dirac and color algebra. We perform our calculation in Feynman gauge.
We express the complete amplitude, fermionic and bosonic contributions, as linear combination
of Feynman integrals belonging to 1, 21 and 107 integral families at three, four and five loops, respectively.

We utilize IBP~identities~\cite{Tkachov:1981wb,Chetyrkin:1981qh} 
and deploy Laporta's algorithm \cite{Laporta:2000dsw} to express all integrals
appearing in the amplitude through a smaller set of integrals 
(for public tools see Refs.~\cite{Lee:2013mka,vonManteuffel:2012np,Smirnov:2019qkx,Maierhoefer:2017hyi,Peraro:2019svx}).
The reduction of the five-loop integrals constitutes the major bottleneck. 
The integral families at five loops contain
up to 12 propagators and 8 irreducible numerators. 
Given that the amplitude needs an integral reduction 
in the top sector with up to 5 scalar products,
this would generate a rich combinatorics when seeding 
the IBP vectors for the construction of IBP equations, 
leading to a huge RAM consumption of several TBs.

In order to perform the IBP reduction, we find beneficial to integrate out the lepton-neutrino loop, 
which corresponds to a massless propagator-like one-loop integral of the form
\begin{multline}
    \int d^d p \frac{p^{\mu_1} \dots p^{\mu_N}}{(-p^2) [-(p-q)^2]} =
    \frac{i \pi^{2-\epsilon}}{(-q^2)^\epsilon}\\
    \times \sum_{i=0}^{[N/2]} 
    f(\epsilon,i,N) \left(\frac{q^2}{2} \right)^i \{[g]^i[q]^{N-2i}\}^{\mu_1 \dots \mu_N},
    \label{eqn:masslessSE}
\end{multline}
where the function $f(\epsilon,i,N)$ is product of Euler's gamma functions (see e.g.~\cite{Smirnov:2012gma})
and the symbol $\{[g]^i[q]^{N-2i}\}^{\mu_1 \dots \mu_N}$
stands for the tensor composed of $i$ metric tensors and $N-2i$ vectors $q$, totally symmetric in its indices.
We rewrite the original five-loop topologies into four-loop ones that have a reduced number
of propagator power indices, 14 instead of 20, where one propagator is now raised to a symbolic power $a_0$. 

\begin{figure}
     \centering
    \begin{tabular}{c}
         \includegraphics[width=0.45\textwidth,clip,trim=2.5cm 0 2.5cm 0]{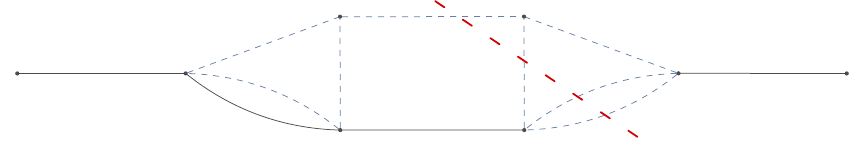}\\
         (a) \\[5pt]
         \includegraphics[width=0.45\textwidth,clip,trim=3cm 0 3cm 0]{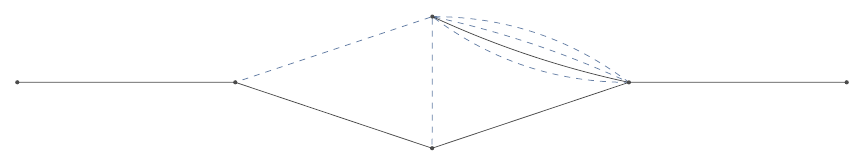}\\
         (b)
    \end{tabular}
    \caption{Example of five-loop Feynman integrals.
    Black and dashed lines represent massive and massless propagators, respectively.
    Integral a) has an imaginary part and we retain its sector during IBP reduction.
    Integral b) has no physical cut so integrals belonging to the same sector can be discarded.}
    \label{fig:mi}
\end{figure}

We work with the IBP reduction program \texttt{Kira}~\cite{Maierhofer:2017gsa,Klappert:2020nbg}
together with the finite field reconstruction library \texttt{FireFly} \cite{Klappert:2019emp,Klappert:2020aqs}. 
In particular \texttt{Kira} supports reductions with symbolic powers. 
First, we perform a reduction of seed integrals with at most two dots and one scalar product.
After identifying the nontrivial sectors in each family, 
we study which sectors contain integrals with a physical cut and therefore
an imaginary part (see for instance Fig.~\ref{fig:mi}a).
Sectors whose integrals are only real valued (see the example in Fig.~\ref{fig:mi}b) are
neglected during the IBP reduction.
We observe that for some families with several massive propagators, such sector selection allows to
eliminate up to 70\% of the nontrivial sectors.

After this step, we proceed with the IBP reduction of the complete set of integrals appearing in the amplitude.
We replace the symbolic power $a_0$ with $(4-d)/2=\epsilon$ in the IBP equations in order to have only one reduction variable instead of two.
Moreover, we take the IBP vectors and eliminate redundant shift operators,
in particular we eliminate most of the operators which would shift the symbolic power. 
When seeding the IBP identities, it is sufficient to have only zero or negative shifts 
of the symbolic power to reduce all necessary integrals.

The list of trivial sectors and symmetries derived for the five-loop topologies
are translated to the four-loop ones. 
The four-loop master integrals are chosen such that no shift of the symbolic propagator power is allowed. 
The translation of four-loop master integrals back to five-loop master integrals is then trivial using Eq.~\eqref{eqn:masslessSE} in the reverse direction. 

For the fermionic contributions, the amplitude is reduced to 1369 master integrals with \texttt{Kira},
which have in the worst case one scalar product, belonging to 48 different integral families.
The complete amplitude is reduced to 8845 master integrals 
which have up to two scalar products belonging to 107 integral families.
Our setup is cross checked with \texttt{FIRE} for ten integral families, where we perform the reduction over a prime field and a fixed value of $d$ with \texttt{FIRE6}~\cite{Smirnov:2019qkx}.

To calculate the master integrals, we leverage the method of solving differential equations numerically (see e.g.\ Refs.~\cite{Pozzorini:2005ff,Liu:2017jxz, Lee:2017qql,Mandal:2018cdj,Czakon:2020vql}). 
In this paper, we follow the strategy outlined in~\cite{Moriello:2019yhu} and
the auxiliary mass flow method~\cite{Liu:2017jxz, Liu:2021wks} which is implemented in the 
\texttt{AMFlow} package~\cite{Liu:2022mfb,Liu:2022chg}. 
For similar approaches see also Refs.~\cite{Armadillo:2022ugh,Hidding:2020ytt,Dubovyk:2022frj,Hidding:2022ycg,Fael:2021kyg,Fael:2022rgm,Fael:2022miw,Fael:2023zqr}.\footnote{Analytic results for the three-loop massive form factors are also given in Refs.~\cite{Blumlein:2023uuq,Blumlein:2019oas}.}
We calculate the five-loop master integrals numerically requiring 40 digits of precision 
using subroutines provided in \texttt{AMFlow}. 
We do not minimize the number of master integrals across different integral families
because it is not crucial for the reduction of the overall runtime for the whole calculation.

The auxiliary mass flow method requires to construct systems of differential equations with respect to
the auxiliary mass $\eta$ which is introduced into certain propagators.
We implement in the framework of \texttt{AMFlow} our own interface to \texttt{Kira} in order to 
perform the IBP reduction with the mapping from five-loop to four-loop topologies as
described above. 
However at this stage of the calculation one needs to consider all non-trivial sectors,
not only those which generate an imaginary part.
The IBP reductions to master integrals are more involved compared to the amplitude reduction
since the additional scale $\eta$ increases the number of master integrals.

\section{Results and discussion}
\label{sec:results}
After the evaluation of the master integrals, we insert their results into the
amplitude and perform the wave function and bottom mass 
renormalization in the on-shell scheme~\cite{Melnikov:2000zc,Chetyrkin:2004mf,Marquard:2016dcn,Marquard:2018rwx},
while we use $\overline{\mathrm{MS}}$ for the strong coupling constant.
The total rate for $b \to u$ decay can be written as
\begin{align}
    \Gamma (B \to X_u \ell \bar \nu_\ell) &= 
    \Gamma_0 
    \left[
    1
    +C_F \sum_{n\ge 1} 
    \left( \frac{\alpha_s}{\pi} \right)^n
    X_n
    \right] 
    \notag \\ &
    +O\left( \frac{\Lambda_\mathrm{QCD}^2}{m_b^2}\right),
\end{align}
where $\Gamma_0 = G_F^2m_b^5 |V_{ub}|^2 A_\mathrm{ew} /(192 \pi^3)$,
$C_F=4/3$ and $\alpha_s \equiv \alpha_s^{(5)}(\mu_s)$ 
is the coupling constant at the renormalization scale $\mu_s$.
$A_\mathrm{ew} = 1.014$ is the leading electroweak correction \cite{Sirlin:1981ie}
and $m_b$ is the on-shell mass of the bottom quark.
The coefficient $X_3$ at order $\alpha_s^3$ can be divided into 
10 color structures:
\begin{align}
    X_3 &= 
    N_L^2 T_F^2 X_{N_L^2} +
    N_H^2 T_F^2 X_{N_H^2} +
    N_H N_L T_F^2 X_{N_H N_L}
    \notag \\ &
    +N_L T_F (C_F X_{N_L C_F}
    + C_A X_{N_L C_A})
    \notag \\ &
    +N_H T_F (C_F X_{N_H C_F}
    + C_A X_{N_H C_A})
    \notag \\ &
    +C_F^2 X_{C_F^2} 
    +C_F C_A X_{C_F C_A}
    + C_A^2 X_{C_A^2},
    \label{eqn:colorstructures}
\end{align}
with $C_F = (N_c^2-1)/(2N_c)$, $C_A = N_c$ and $T_F = 1/2$ for 
an $SU(N_c)$ gauge group.
Here $N_L = 4$ is the number of massless quarks and $N_H= 1$ labels the $b$-quark loop.
The first seven color structures are the fermionic contributions while 
the last three stem from diagrams where only gluons are exchanged.

We estimate the precision of our result from the numerical pole cancellations of the renormalized decay rate. 
We have analytic expressions for the bare amplitude up to order $\alpha_s$, while 
at $O(\alpha_s^2)$ the amplitude is obtained via numerical evaluation of the master integrals with 80 digits of precision. 
We observe that in $X_3$ the $\epsilon^{-3}, \epsilon^{-2}$ and $\epsilon^{-1}$ poles cancel with more than 37, 35 and 33 digits, respectively.
Extrapolating those numbers to the finite terms, we expect that our results are correct up to 30 digits.

\begin{table}[tbh]
\begin{ruledtabular}
\begin{tabular}{rrrr}
     & \text{This work} & \text{Ref.~\cite{Fael:2020tow}} & \text{Difference} \\
    \hline
    $T_F^2 N_L^2$   & -6.9195 & -6.34 (42) & 8.3\% \\   
    $T_F^2 N_H^2$   & $-1.8768 \times 10^{-2}$ & $-1.97 \,(42) \times 10^{-2}$& 5.0\% \\   
    $T_F^2 N_H N_L$ & $-1.2881 \times 10^{-2}$ & $-1.1 \, (1.1) \times 10^{-2}$  & 12\%  \\   
    $C_F T_F N_L$     & -7.1876 & $-5.65\,(55) $ & 22\%  \\   
    $C_A T_F N_L$   & 42.717 & $ 39.7\, (2.1)$ & 7\% \\   
    $C_F T_F N_H$     & 2.1098  & $2.056 \, (64)$& 2.5\% \\   
    $C_A T_F N_H$   & -0.45059& $-0.449 \,(18)$& 0.4\% \\   
\end{tabular}
\end{ruledtabular}
\caption{The first five digits of the color structure coefficients
of $X_3$ in Eq.~\eqref{eqn:colorstructures} at a renormalization
scale $\mu_s = m_b$.
The third and fourth column report the value from Ref.~\cite{Fael:2020tow}
and the relative difference, respectively.}
\label{tab:results}
\end{table}

We present in Tab.~\ref{tab:results}
compact results for the first seven color factors at the renormalization
scale $\mu_s = m_b$, reporting in the second column the first five significant digits. 
In the third column, we compare our results with Ref.~\cite{Fael:2020tow}
where we use the asymptotic expansion up to $\delta^{12}$ for the central value and estimate the uncertainty from the difference between the $\delta^{11}$ and $\delta^{12}$ expansion, 
multiplied by a security factor of five.
We observe overall a good agreement within the 
uncertainties, except for the color structure $X_{N_L C_F}$ where the
deviation is larger than the uncertainty based on the asymptotic series convergence.

As a cross check, we compare our finding with Ref.~\cite{Chen:2023dsi} which presents analytic 
expressions for the leading-color contributions to $X_3$. 
After taking the large-$N_c$ limit, our results proportional to $N_L^2$ and $N_L$ agree with Eq.~(13) of Ref.~\cite{Chen:2023dsi} with more than 30 digits.

We also update the prediction for the third order correction
in the on-shell scheme:
\begin{widetext}
\begin{equation}
    \Gamma(B \to X_u \ell \bar \nu_\ell) = \Gamma_0 
     \left[
    1-2.413 \frac{\alpha_s}{\pi}
    - 21.3\left(\frac{\alpha_s}{\pi}\right)^2
    -267.8 \,(2.7)\left(\frac{\alpha_s}{\pi}\right)^3
    \right],
\end{equation}
\end{widetext}
where $X_3 = -200.9 \pm 2.0$.
The value at $O(\alpha_s^3)$ is obtained by summing our fermionic contributions
and the analytic expression for the 
bosonic contribution in the large-$N_c$ limit from Ref.~\cite{Chen:2023dsi}.
Moreover, we add the subleading color terms which result from the calculation
in Ref.~\cite{Fael:2020tow}. 
The quoted uncertainty arises from the massless extrapolation
and it is estimated as in Tab.~\ref{tab:results}.
The uncertainty is reduced by a factor of four compared to Ref.~\cite{Chen:2023dsi} and a factor of ten with respect to Ref.~\cite{Fael:2020tow}.

\section{Conclusions}
\label{sec:conc}
We presented the fermionic contributions to the
decay rate of $b \to u \ell \bar \nu_\ell$ at order $\alpha_s^3$.
Our calculation is based on IBP reductions of Feynman integrals with a symbolic 
propagator power and numerical evaluation of master integrals via the 
auxiliary mass flow method.
We estimate that our results have an accuracy of at least thirty digits.

The calculation of the missing three color structures coming from 
the bosonic contributions is ongoing.
The method described in this paper can also be applied to the calculation 
of the finite charm-mass effects, although requiring additional computer resources due to the new scale $m_c/m_b$ appearing in the Feynman integrals.

The decay rate $\Gamma(B \to X_u \ell \bar \nu_\ell)$ is 
an important ingredient in the normalization of radiative and rare semileptonic
decays of $B$ meson and can be employed to reduce the current theoretical 
uncertainty on the phase-space ratio $C$.

\begin{acknowledgments}
We thank Joshua Davies and Xiao Liu for valuable advice in the usage of FORM and AMFlow, respectively. 
After acceptance of our manuscript, the authors of Ref.~\cite{Chen:2023osm} 
made available their unpublished results for the coefficients of $N_L$ and $N_L^2$ in Eq.~\eqref{eqn:colorstructures}, 
which agree well with our results in the Table~\ref{tab:results}.
The work of M.F.\ is supported by the European Union’s Horizon 2020 
research and innovation program under the Marie-Sk\l odowska-Curie 
Grant Agreement No.\ 101065445 - PHOBIDE.

\end{acknowledgments}

\bibliographystyle{apsrev4-1}
\bibliography{biblio}

\end{document}